\documentclass{cta-author}

{}
{}
{}
\usepackage[algo2e]{algorithm2e}
\usepackage{algorithm,algpseudocode}
\usepackage{tabularx}
\usepackage{diagbox}
\usepackage{multirow}
\usepackage{makecell}
\newcommand{\thickhline}{%
	\noalign {\ifnum 0=`}\fi \hrule height 1.5pt
	\futurelet \reserved@a \@xhline
}
\begin{document}

\supertitle{Submission Template for IET Research Journal Papers}

\title{Playback Experience Driven Cross Layer Optimization of APP, Transport and MAC Layer for Video Clients over LTE System}
\author{\au{Xinyu Huang$^{1}$}, \au{Lijun He$^{2\corr}$}}

\address{\add{1}{School of Information and Communications Engineering, Xi'an Jiaotong University, Xianning Road, Xi'an 710049, China}
	\add{2}{School of Information and Communications Engineering, Xi'an Jiaotong University, Xianning Road, Xi'an 710049, China}\email{jzb2016125@mail.xjtu.edu.cn}}

\begin{abstract}
In traditional communication system, information of APP (Application) layer, transport layer and MAC (Media Access Control) layer has not been fully interacted, which inevitably leads to inconsistencies among TCP congestion state, clients' requirements, and resource allocation. To solve the problem, we propose a joint optimization framework, which consists of APP layer, transport layer and MAC layer, to improve the video clients' playback experience and system throughput. First, a client requirement aware autonomous packet drop strategy, based on packet importance, channel condition and playback status, is developed to decrease the network load and the probability of rebuffering events. Further, TCP (Transmission Control Protocol) state aware downlink and uplink resource allocation schemes are proposed to achieve smooth video transmission and steady ACK (Acknowledgement) feedback respectively. For downlink scheme, maximum transmission capacity requirement for each client is calculated based on feedback ACK information from transport layer to avoid allocating excessive resource to the client, whose ACK feedback is blocked due to bad uplink channel condition. For uplink scheme, information of RTO (Retransmission Timeout) and TCP congestion window are utilized to indicate ACK scheduling priority. The simulation results show that our algorithm can significantly improve the system throughput and the clients' playback continuity with acceptable video quality.
\end{abstract}

\maketitle

\section{Introduction}\label{introduction}
With the development of video compression and wireless communication technologies, video service has become the most effective way to convey information. The resulting large amount of video data brings enormous pressure on the wireless transmission. Online VoD (Video on Demand) services enable clients to watch video content such as client generated videos, movies, music videos, and live streams \cite{Juluri_2016}. The reliable transport protocol such as TCP is often employed in online VoD services, which can guarantee successful transmission without packet loss \cite{He_2018}. However, in the traditional communication system, the layers are independent of each other such that their information cannot be detected and interacted among them, which inevitably leads to low resource efficiency\cite{Pacifico_2009,Zulhasnine_2010}. Moreover, TCP protocol blindly guarantees the successful transmission of each video packet even if the packet contributes little to the received video quality\cite{Tanigawa_2017}. Therefore, analyzing the redundant information in the video content to reduce the amount of video data and optimizing the parameters of each layer which involves the video transmission, are two effective ways to provide satisfactory experience to video clients.

As the current mainstream reliable transmission protocol, there have been many researches on the transmission optimization of the transport layer. Due to the time-varying characteristic of the  wireless network state, the transmission capability of the wireless channel is also time-varying. When the transmission status of TCP cannot respond to the change of the wireless network state in time, network congestion will occur. Most of the existing works to optimize the transport layer focus on the optimization of the TCP congestion mechanism. The authors in \cite{Pawale_2017} utilized the sending window and RTT (Round-Trip Time) to update the slow start threshold, which in turn reduced the number of times that the New Reno protocol mistakenly reduced the TCP window size. However, the system throughput was relatively lower during the initial phase of data transmission. The authors in \cite{Cheng_2017} combined the bandwidth estimation strategy and the rate estimation strategy to calculate the effective bandwidth, classified the network congestion according to the throughput, and estimated the congestion degree, thereby improving the performance of TCP in the LTE (Long Term Evolution) network. However, the fairness of the algorithm was poor as the packet loss rate increased. The authors in \cite{Evg_2015} proposed a low-complexity unequal packet loss protection and rate control algorithms for scalable video coding based on the 3-D discrete wavelet transform, which provide a feasible solution for real-time automotive surveillance applications. In \cite{Singh_2018}, the authors adjusted the transmission window size by calculating the estimated bandwidth and dynamic congestion control factor, which improved the system throughput and algorithm fairness. However, the accuracy of the estimated bandwidth was low. The authors in \cite{Dalal_2015} estimated the congestion degree by using the statistics of RTT, thus avoiding unnecessary shrinkage of lost window, while the fairness cannot be guaranteed among clients. In \cite{Ameur_2017}, the authors optimized the bandwidth estimate and slow start threshold to improve the QoE (Quality of Experience) of clients and the QoS (Quality of Service) of network, but the sending window cannot respond to the wireless channel variation in time when the wireless channel varied frequently. The authors in \cite{Han_2018} employed the deep learning method to distinguish network congestion and random packet loss, to solve the problem of TCP performance degradation caused by network congestion. Nevertheless, this cannot avoid the decrease of the window size due to the packet loss. Overall, the above work advances the optimization of the TCP congestion mechanism. However, when the wireless network conditions become worse and the network load becomes larger, even the perfect congestion mechanism cannot solve the huge delay caused by the wireless network, which eventually reduces the QoE of video clients \cite{Li_2018,He_2014}. The scalable video coding \cite{Schwarz_2007} and the storage of different copies of the same video compressed at different quantization steps \cite{Wiegand_2003} are two effective methods to transmit the videos for the client whose channel state is not well enough to support the high-rate video transmission. However, these two methods cannot adjust the size of video according to the change of the channel state continuously, because both coding schemes cannot produce the discrete values of video size. In fact, video data has different characteristics compared with general data, and complex dependence exists among the video packets in one video sequence. This relationship enables video data to obtain satisfied video quality even in the case of packet loss. Based on the playback information of the client and the transport layer congestion information, some unimportant video packets should be reasonably dropped before putting them into the TCP sending window. This can not only reduce the probability of the sudden drop of the TCP window caused by the loss of some unimportant video packets but also decrease the wireless network load. Besides, as another mainstream communication protocol, UDP (User Datagram Protocol) cannot provide any congestion control mechanism but can bring less delay, which is commonly used in real-time video and audio streaming services \cite{Iqbal_2010}. For wireless real-time multimedia applications, the authors in \cite{Larzon_1999} and \cite{Lam_2004} proposed the UDP-Lite and UDP-Liter protocol to solve the problem of excessive loss rate suffered at the receiver. In addition, to guarantee the reliable transmission, the authors in \cite{Wang_2017} proposed a reliable UDP-based transmission protocol over SDN (Software Defined Network), which can significantly reduce the overhead of TCP traffic.

The smooth transmission of video data cannot be achieved only by optimizing the single layer, i.e. transport layer, because the layers of the complex communication system interact with each other. Currently, there is a lot of work devoted to optimizing the layers of the communication network to improve system throughput. The most classic one is the resource allocation algorithm for the MAC layer. The authors in \cite{RR} provided a fair resource allocation at the MAC layer without considering any QoS requirement, where the resource was just allocated in a round-robin mode. The two state-of-art works in \cite{Maxci,PF} modified the traditional resource allocations to maximize the system throughput and guarantee the proportional fairness among the clients, respectively. However, the requirement of the APP layer has not been considered in all the above three schemes, which made their performance limited and not suitable to be applied in video transmission. Focusing on the APP requirements, the authors in \cite{MLWDF} considered the packet urgency to allocate the wireless resource to guarantee the smooth video transmission, but it ignored the client buffer status that made the estimated packet urgency inaccurate. To handle this obstacle, the authors in \cite{Iturralde_2011} employed the client's QoS requirement and buffer size to better allocate wireless resource, but it performed badly when the initial delay is high. Furthermore, the authors in \cite{Li_2016}, provided a delay priority scheduling scheme, which improved the QoS performance of real-time communication at the expense of acceptable QoS performance under non-real-time communication. The authors in \cite{Ramesh_2018} employed the Kalman filter algorithm to estimate the number of RBs (Resource Blocks) for uplink ACK transmission at the MAC layer, but the method of downlink transmission needed to be optimized to avoid allocating excessive resource to the client. In \cite{Kayali_2017}, the authors considered the QoS requirement of the network and the buffer status of the client, which reduced the packet loss caused by the buffer overflow and improved the throughput of the system, nevertheless, the fairness among clients was poor. 

In addition to the optimization of the MAC layer, there is still much research work to jointly optimize the MAC layer and other layers. In \cite{Kwon_2005}, the authors proposed a cross layer optimization algorithm combining the MAC layer and the physical layer. The uplink control channel was employed to exchange physical channel information and ACK information, which improved system throughput under specified QoS, but the timeout retransmission strategy didn't consider service delay bound and channel quality. Similarly, the authors in \cite{Karachontzitis_2011} also proposed a cross layer optimization algorithm for MAC layer and physical layer. Specifically, video packets were scheduled according to packet importance in the MAC layer, and a linear SDMA (Space-Division Multiple Access) algorithm was proposed at the physical layer, which guaranteed the client's QoS, improved the performance of the system and reduced the complexity of the algorithm. Nevertheless, the system throughput cannot be improved when the number of clients in the cell increased to a high level. In \cite{Wang_2012}, the authors proposed a cross layer optimization algorithm for the physical layer and the APP layer, which combined CSI (Channel State Information) and RD (Rate-Distortion) information to allocate RBs under the specified system throughput and PSNR (Peak Signal to Noise Ratio), but it did not consider the playback status of client which may decrease the QoE of clients. In \cite{Qadir_2016}, the authors proposed a cross-layer optimization algorithm for the APP layer and the network layer. A combination of the rate adaptation at the APP layer and QoE-aware admission control at the network layer was the core of algorithm, which reduced the average delay and packet loss rate. In \cite{Nasimi_2017}, the authors proposed a cross layer optimization algorithm for the APP layer, the physical layer and the MAC layer, to maximize the utilization of network resource and client's QoE. The above work has greatly improved the data transmission of the wireless network, but the joint optimization of the APP layer, the transport layer and the MAC layer has not been involved. The client's real playback information, ACK feedback condition, the TCP congestion state and resource allocation at the MAC layer are interacted on each other and influence the system throughput as well as the clients' playback experience. To provide satisfactory playback experience for video clients, it is important to deeply explore the relationship between the APP layer, the transport layer and the MAC layer.

Generally, there are some problems to be addressed for video transmission over TCP: (1) The information of transport layer and MAC layer has not been fully interacted and utilized; (2) The video content information has not been investigated intensively and some redundant information still exists. Inspired by the above prior works, we propose a joint optimization framework, which consists of the APP layer, the transport layer and the MAC layer, to improve the performance of video transmission in LTE system. Our work is novel in the following aspects.

\begin{itemize}
	\item A new joint optimization framework of the APP layer, the transport layer and the MAC layer for video transmission
	
	To overcome the independence among the layers of traditional communication system, we built a new framework which can support the information delivery among the App layer, the transport layer and the MAC layer, such that the optimization of the TCP and the MAC layer can be achieved by employing the information from other layers.
	
	\item QoE requirement aware autonomous packet drop strategy
	
	Without any change of TCP traditional protocol, we develop an APP layer QoE requirement aware autonomous packet drop strategy to decrease the network load as well as transmit the packets, which are really important to the received video quality. In this strategy, the playback status of client and video coding information are jointly used to adjust the amount of video packets to be transmitted.
	
	\item TCP state aware downlink and uplink resource allocation scheme
	
	Continuous transmission requires steady downlink packet transmission and timely ACK feedback. To guarantee continuous transmission, we first employ the ACK feedback information from transport layer to calculate the transmission capacity requirement for downlink resource allocation to prevent from allocating excessive resources to the client with bad uplink channel state. We also predict the urgency of ACK based on TCP sending information for uplink resource allocation at MAC layer.
	
\end{itemize}

The rest of the paper is organized as follows. Section \ref{system} introduces the proposed system model. Section \ref{APD} presents a APP layer QoE requirement aware autonomous drop strategy at the transport layer. Section \ref{congestion_aware} proposes the TCP congestion state aware resource allocation schemes for uplink and downlink video transmission respectively. Section \ref{result} provides numerical results. Finally we draw the conclusion of the paper in Section \ref{conclusion}. 

\section{System Model}\label{system}

\begin{figure*}[!h]
	\centering
	\includegraphics[width=16cm]{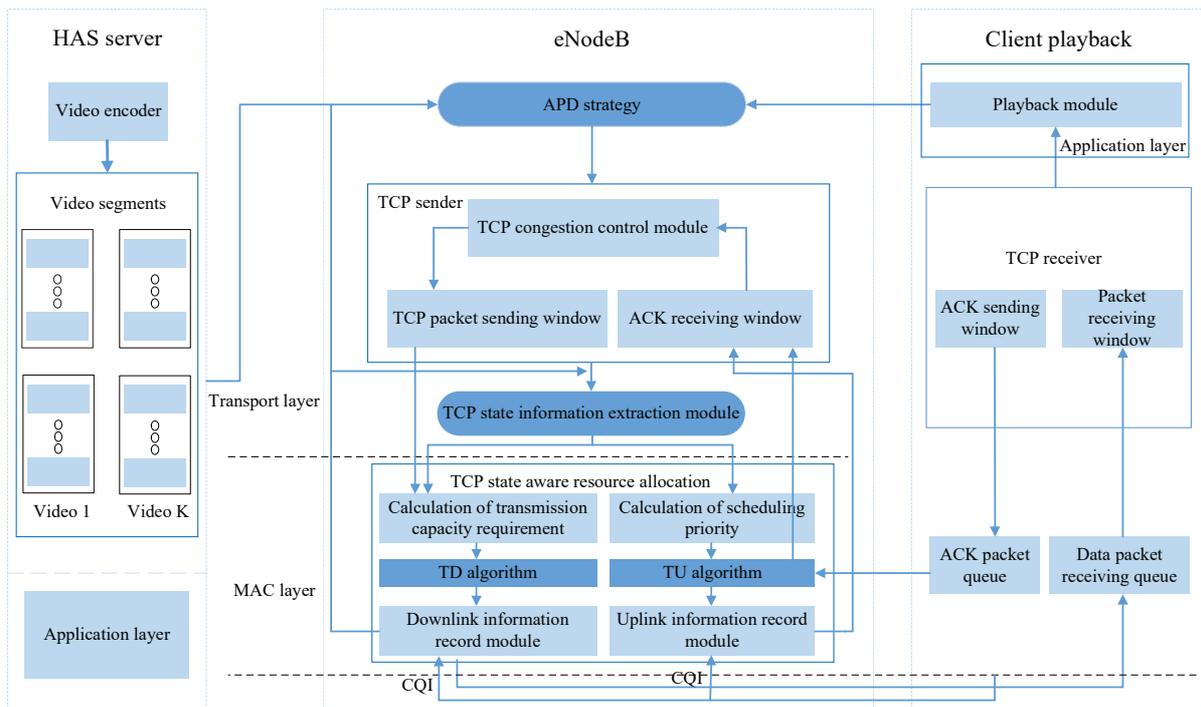}
	\caption{System framework}
	\label{fig:kuangtu}
\end{figure*}

In this study, we propose an optimization framework, which consists of the APP layer, the transport layer and the MAC layer, to jointly improve the QoE of the clients. The major system framework is shown in Fig. \ref{fig:kuangtu}. 

The whole system framework contains three parts: HAS (HTTP Adaptive Streaming) server, eNodeB (Evolved NodeB) and playback module of client. Specifically, HAS server is used to provide the video sequences requested by the clients\cite{Sto_2011}. To guarantee the efficient transmission and continuous playback, each video sequence is pre-encoded into a series of video segments by H.264/AVC (Advanced Video Coding) encoder offline. All the segments stored at a HTTP (HyperText Transfer Protocol) media server is attached to eNodeB through a lossless wired network, which are delivered to the TCP sender for data transmission. At transport layer of eNodeB, an APP layer QoE requirement aware APD (Autonomous Packet Drop) strategy is developed to keep a balance between the client's QoE and wireless network condition based on the information of client and the MAC layer. Besides, a TCP state information extraction module is added at the transport layer to collect and send TCP state information to the resource scheduling module at the MAC layer. Based on the extracted TCP state information, we calculate the upper bound of transmission capacity of the MAC layer at the module of calculation of transmission capacity requirement. Then a feedback ACK information based downlink resource allocation scheme to avoid allocating excessive resource to the client is proposed. Besides, to improve the uplink feedback of ACK, we utilize the ACK information collected by uplink information record module from the TCP congestion information  to determine the scheduling priority of ACK packet.

\section{An APP Layer QoE Requirement Aware APD Strategy at transport layer}
\label{APD}
\begin{figure}[!h]
	\centering
	\includegraphics[width=8.5cm]{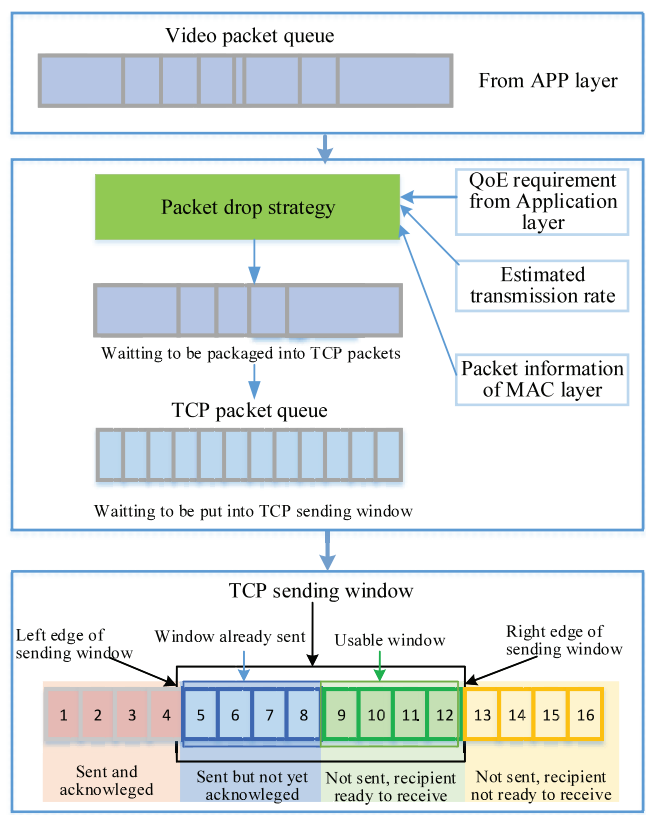}
	\caption{The design diagram of the packet drop module}
	\label{fig:diubao}
\end{figure}

Due to the retransmission mechanism, the traditional TCP can guarantee the successful transmission of all the video packets. The loss of each packet will cause retransmission and degradation of congestion window size even if it is very small. In fact, different video packets will bring different effects on the received video quality. For the packets with small size and little contribution to the received video quality, we can drop them autonomously according to the current TCP congestion state before packaging the original video packets into TCP packets. In one hand, this can decrease the bits to be transmitted and the packet loss possibility, which can provide higher opportunity to keep more steady congestion window state. Moreover, since the loss of these packets has no or only a little impact on the received video quality, the requirement of the received video quality can also be satisfied. In the other hand, it is not necessary to change the TCP protocol. We just need to add an autonomous packet drop module before packaging the original video packets into TCP packets. The design diagram of the packet drop module is shown in Fig. \ref{fig:diubao}.

Firstly, all the video packets from the APP layer will be determined whether to be packaged into TCP packet or not based on the information extracted from the APP layer of client and the estimated transmission rate. Then, the left packets will be packaged into TCP packets and put into TCP send window. Note that, once the packet is packaged into a TCP packet or put into TCP send window, it cannot be dropped but be transmitted to client. Next, we will describe how to determine the packet importance and the packet drop strategy.

(1)  Packet Importance Analysis

In our opinion, the packet importance can be regarded as the potential degradation of the video quality caused by the loss of the packet. When the packet is lost, the pixels of the corresponding location in the latest decoded reference frame will be used to cover the pixels of the lost packet in the current frame. We employ the information of motion vectors, coding modes and the reference relationship as described in our previous work \cite{pre_2009} to calculate the priority of each packet.

Let $U_{k,f,m}$ indicate the importance of packet $m$ of frame $f$ of client $k$. For each packet, $U_{k,f,m}$ can be described as
\begin{equation}
U_{k,f,m}=D_{k,f,m}^{cur}+D_{k,f,m}^{ref}
\end{equation}
where $D_{k,f,m}^{cur}$ and $D_{k,f,m}^{ref}$ are the relative distortion caused by the reconstructed packet and the following packets dependent on it respectively. Let $N_x$ and $N_y$ be the numbers of intra-coded and inter-coded MBs (Macro Blocks) of packet $m$ of frame $f$, respectively. $D_{k,f,m}^{cur}$  can be estimated based on the coding parameters,
\begin{equation}
D_{k,f,m}^{cur}=\sum_{\varsigma=1}^{N_x}\omega_{\varsigma}+\sum_{\varsigma=1}^{N_x}(\omega_{\varsigma}\sum_{\nu=1}^{N_y}Q_{\varsigma,\nu})
\end{equation}
where $\omega_{\varsigma}$ represents the prediction weight of MB $\varsigma$.
Since the intra-coded and inter-coded prediction modes are supported by H.264 coding standard. The selection of the prediction modes for an MB depends on the degree of similarity between the current MB and that in the reference frame. If a MB uses the intra-coded prediction mode, it means that there are no MB matches in the reference frame. Therefore, such an MB should be given a larger $\omega_{\varsigma}$. Otherwise, a MB would use the inter-coded prediction mode, and hence, a smaller $\omega_{\varsigma}$ will be allocated to it. Different partition modes also needs different $\omega_{\varsigma}$. For intra-coded prediction mode, the 16*16 partition mode should be given a lower $\omega_{\varsigma}$ than the one using the 4*4 partition mode. $Q_{\varsigma,\nu}$ is the relative motion intensity of block $\nu$ in the inter-coded MB $\varsigma$, which can be expressed by
\begin{equation}
Q_{\varsigma,\nu}=\sqrt{(\frac{MV_{\varsigma,\nu}^x}{W})^2+(\frac{MV_{\varsigma,\nu}^y}{H})^2}
\end{equation}
where $(MV_{\varsigma,\nu}^x,MV_{\varsigma,\nu}^y)$ are the motion vector of block $\nu$ in the inter-coded MB $\varsigma$. $W$ and $H$ are the corresponding width and height. $D_{k,f,m}^{ref}$ can be regarded as the sum distortion of the packets dependent on the current one, which can be estimated by $D_{k,f,m}^{ref}=\beta(k,f,m)D_{k,f,m}^{cur}$. $\beta(k,f,m)$ is the percentage of the number of the pixels referred in packet $m$ of frame $f$. Since the packet index can indicate each individual video packet, the frame index, $f$, is omitted in the following discussion.

(2) Packet Drop Strategy

To determine which packet should be dropped, first we need to predict the transmission capacity of client $k$ to guarantee continuous playback, $CP_k$. Let $t_k$ denote the continuous playback time that the unplayed frames of the completely received segments of client $k$ can support. Let $TR_k$ be the estimated transmission rate, which can be calculated by
\begin{equation}
TR_k=\Upsilon\widehat{TR_k}+(1-\Upsilon)\overline{TR_k}
\end{equation}
where $\Upsilon\in(0,1)$, $\widehat{TR_k}$ is the real transmission rate of client $k$ for the previous time slot and $\overline{TR_k}$ is the average transmission rate of client $k$ for previous $GT$ time slots. Then $CP_k$ can be obtained by
\begin{equation}
CP_k=TR_k\times (t_k+g_k)
\end{equation}
where $g_k$ is the guard time interval to decrease the effect of inaccurate estimation of the transmission rate. It is easy to see that larger $g_k$  means fewer packets to be dropped. If continuous playback can be guaranteed, we can have
\begin{equation}
CP_k=S_{k}^{mac}+S_{k}^{send}+\lambda_k\cdot S_{k}^{vq}
\end{equation}
where $S_{k}^{mac}$, $S_{k}^{send}$ and $S_{k}^{vq}$ are the sum size of the packets waiting to be scheduled at MAC layer, TCP send window and video packet queue in Fig.\ref{fig:diubao}. $\lambda_k$ is the proportionality factor of packet drop, which can be computed by
\begin{equation}
\lambda_k=\frac{CP_k-(S_{k}^{mac}+S_{k}^{send})}{S_{k}^{vq}}
\end{equation}

If $\lambda_k\geq 1$, it implies that the wireless resource can afford the transmission of all the video packets so that it is not necessary to drop any video packets. If $\lambda_k<1$, it means that he transmission capacity is not enough, it is necessary to drop some packets. Let $S_{k}^{drop}$ denote the sum size of the packets which can be dropped such that
\begin{equation}
S_{k}^{drop}=(1-\lambda_k)S_{k}^{vq}
\end{equation}

Let $MP_k$ be the number of the packets of client $k$. Since the objective of our proposed packet drop strategy is to minimize the quality degradation caused by the packet loss, the packet drop problem can be formulated as follows:
\begin{equation}
\begin{split}
&\min\sum_{m=1}^{MP_k}U_{k,m}\rho_{k,m}\\
&s.t.\left\{
\begin{aligned}
&\sum_{m=1}^{MP_k}R_{k,m}\rho_{k,m}\geq S_{k}^{drop}\\
&\rho_{k,m}\in\{0,1\}\\
\end{aligned}
\right.
\end{split}
\end{equation}
where $\rho_{k,m}$ is an integer binary variable which indicates whether packet $m$ of client $k$ is dropped or not. $R_{k,m}$ is the size of packet $m$ of client $k$. $\rho_{k,m}=1$ means packet $m$ of client $k$ will be dropped, otherwise $\rho_{k,m}=0$. The constraint implies that we need to drop enough video packets to guarantee continuous playback. The above problem is a 0-1 knapsack problem which is easy to be solved.

\section{TCP Congestion State Aware Resource Allocation for Uplink and Downlink Video Transmission}\label{congestion_aware}
TCP congestion mechanism can adjust the TCP sending rate based on the feedback ACK. Since the feedback ACK is mostly influenced by the wireless network state, TCP congestion window can be adaptive to the time-varying channel state to some extent \cite{Fou_2008}. However, sometimes the wireless channel state varies so frequently that the adjustment of TCP congestion cannot keep up with the variation. Once retransmission event occurs, the degraded congestion window cannot be recovered immediately even though the current wireless channel state becomes good enough to support large TCP congestion window size. The resource allocation scheme at MAC layer may still schedule a large amount of packets for the client without considering its TCP congestion state information. If the latest retransmission is caused by blocked ACK feedback, more retransmissions will occur. To handle this obstacle, TCP state information extraction module is added to collect the TCP state information. We utilize the TCP state information of transport layer to obtain maximum transmission capacity for downlink resource allocation and ACK packet scheduling priority for uplink transmission, to improve the system throughput.

\subsection{TCP Congestion State Aware Resource Allocation for Downlink Video Transmission}\label{downlink}
Without considering the information of the transport layer, the traditional resource allocation schemes may allocate the limited resource to afford the transmission of the packets, whose ACKs cannot be received in time, especially in a scenario with bad or unstable uplink channel state. To avoid the waste of resource and retransmissions, an uplink feedback ACK information-based downlink resource allocation scheme is proposed to achieve more reasonable resource allocation.

(1) Calculation of the Downlink Transmission Capacity Requirement at the MAC Layer of Client $k$

TCP state information extraction module mainly obtains the information of the receiving time of each packet and ACK. The receiving time of each packet and ACK is employed to calculate the the ACK feedback rate. Based on the header size of various layers (i.e. TCP, IP, PDCP) and ACK feedback rate, the transmission capacity of each client can be determined.

It is easy to understand that smooth transmission can be achieved under the condition that the corresponding ACK can be received in time for each TCP packet, which has been already sent. Therefore, we need to first estimate the feedback rate of ACK. Let $T_{k,i}$ denote the estimated transmission time of ACK $i$ of client $k$, which can be computed by
\begin{equation}\label{T_k,i}
T_{k,i}=Re_{k,i}-St_{k,i}
\end{equation}
where $St_{k,i}$ is the receiving time of the accepted TCP packet $i$ of client $k$. To be specific, this parameter is the feedback time of ACK, which is from the last sub-packet of the TCP packet $i$ in the MAC layer of eNodeB. $Re_{k,i}$ denotes the receiving time of the ACK $i$ of client $k$ at TCP sender.

Let $RT_k$ denote the average feedback rate of $G_k$ $(1\leq k\leq K)$ ACKs, which can be obtained by the combination of $A$ (the size of ACK) and $T_{k,i}$ as follows.
\begin{equation}\label{R_ack_k}
RT_k=(\sum_{i=1}^{G_k}\frac{A}{T_{k,i}})/G_k
\end{equation}

On the basis of average feedback rate $RT_k$, then the requirements of downlink transmission rate of client $k$ could be depicted as
\begin{equation}\label{R_TCP}
DR_k=RT_k/\alpha
\end{equation}
where $\alpha$ is the ratio of downlink rate of client $k$. It can be calculated by the ratio of the size of ACK and the TCP packet ,$P$, which can be described as $\alpha=A/P$.

Next, we need to obtain the transmission capacity required by client $k$ in MAC layer. When the TCP packet of client $k$ arrives at the RLC (Radio Link Control) layer, the header of TCP, IP (Internet Protocol) and PDCP (Packet Data Convergence Protocol) will be added to the TCP packet, which means that the header sizes of these layers should be added into $P$. Thus, at MAC layer, the size of one TCP packet becomes to 
\begin{equation}\label{S_pdcp}
B_k=S_{headtcp}+S_{headip}+S_{headpdcp}+P
\end{equation}
where $S_{headtcp},S_{headip},S_{headpdcp}$ denote the protocol header size of TCP, IP and PDCP, respectively. Assuming that once the packet arrives at the transmission queue of MAC layer, it will be divided into $Nb_k$ sub-packets. Further, the protocol header size of RLC and MAC is $S_{headmac}+S_{headrlc}$, then the requirement of transmission rate of client $k$ to guarantee fluent transmission of ACK at MAC layer can be expressed as
\begin{equation}\label{MAC_shangxian}
M_k=DR_k\frac{B_k+Nb_k(S_{headmac}+S_{headrlc})}{P}
\end{equation}

To estimate $Nb_k$ precisely, we utilize the average CQI (Channel Quality Index) values of client $k$ in the current scheduling period, $CQI_k$, and the average number of MAC PDU of previous $w$ scheduling periods, $\widetilde{Nb_k}$, then we can get
\begin{equation}\label{N_k}
Nb_k=\frac{\widetilde{Nb_k}\times F_k}{CQI_k}
\end{equation}
where $F_k$ is the average CQI values of client $k$ of latest $w$ scheduling periods. Let $TTI$ denote the duration of one scheduling period of the LTE system. Finally, the transmission capacity required by client $k$ at MAC layer in the current scheduling period can be computed by
\begin{equation}\label{chuanshu_MAC}
C_k=M_k\cdot TTI
\end{equation}

(2) Problem Formulation

The object of downlink transmission is to maximize the total system throughput while satisfying the transmission capacity requirements of all the clients. Let $\Omega$ be the set of the total available RBs, $\Omega=\{1,\cdots,N\}$ and $\mathcal{X}$ be the set of clients waiting to be scheduled, $\mathcal{X}=\{1,\cdots,K\}$. We denote $a_{k,n}$ as the optimization variable, which means whether RB $n$ is assigned to client $k$. Each RB should be assigned to one client and only one client at each scheduling period, thus we can have
\begin{equation}
\sum_{k=1}^{K}a_{k,n}=1,a_{k,n}\in\{0,1\},n\in\Omega,k\in \mathcal{X}
\end{equation}
where $a_{k,n}=1$ represents that RB $n$ is assigned to client $k$ and $a_{k,n}=0$ otherwise.

Let $b_{k,j}$ denote the optimization variable indicating whether client $k$ use MCS (Modulation and Coding Scheme) $j$. All RBs assigned to the same client should employ the same MCS, which means
\begin{equation}
\sum_{j=1}^{q(N_k)}b_{k,j}=1,b_{k,j}\in \{0,1\},k\in \mathcal{X}
\end{equation}
where $b_{k,j}=1$ means client $k$ employs MCS $j$ and $b_{k,j}=0$ otherwise. $N_k$ is the set of RBs assigned to client $k$ and $q(N_k)$ is the maximum MCS that client $k$ can support according to $N_k$.

To guarantee the continuous transmission of data, ACK needs to be received in time. In other words, the transmission capacity of client $k$ should not be smaller than the transmission capacity requirement $C_k$. Hence, the capacity of the assigned RBs can not be greater than the sum size of all the packets in client $k$'s MAC queue, which is depicted as
\begin{equation}\label{lambda}
J_k\geq \sum_{n\in\Omega}a_{k,n}\sum_{j=1}^{q(N_k)}b_{k.j}r_{k,j}\geq C_k,k\in \mathcal{X}
\end{equation}
where $J_k$ is the sum size of the packets in the MAC queue of client $k$ and $r_{k,j}$ is the transmission capacity of one RB when client $k$ uses the MCS $j$. Given MCS $j$, $r_{k,j}$ can be determined by
the Table 7.1.7.2.1-1 in the technical specification\cite{3gpp}. Based on the above discussion, our objective can be formulated to maximize the total system throughput while satisfying the transmission capacity requirements of all the clients:
\begin{equation}\label{mubiao}
\begin{split}
&\mathbf{P}: max\sum_{k=1}^{K}\sum_{n\in\Omega}a_{k,n}\sum_{j=1}^{q(N_k)}b_{k.j}\times r_{k,j}\\
&s.t. \left\{
\begin{aligned}
&\sum_{k=1}^{K}a_{k,n}=1,a_{k,n}\in\{0,1\},n\in\Omega,k\in \mathcal{X}\\
&\sum_{j=1}^{q(N_k)}b_{k,j}=1,b_{k,j}\in \{0,1\},k\in \mathcal{X}\\
&J_k\geq\sum_{n\in \Omega}a_{k,n}\sum_{j=1}^{q(N_k)}b_{k,j}r_{k,j}\geq C_k,k\in \mathcal{X}
\end{aligned}
\right.
\end{split}
\end{equation}

(3) TD (TCP State Aware for Downlink Resource Allocation) algorithm

TD algorithm mainly consists of three steps: (1) Calculate the transmission capacity requirements of all the clients based on the information of feedback ACK. (2) Perform the RB assignment and MCS optimization for all the clients whose transmission capacity have not been satisfied. (3) If there are RBs left, perform the RB assignment for all the clients based on the CQI value. The detailed process of our proposed TD algorithm is depicted in Algorithm \ref{downlink}.

\begin{algorithm}[H]
	\caption{TD algorithm}
	\label{downlink}
	\KwIn{ $Re_{k,i}$, $St_{k,i}$, $A$, $P$, $S_{headtcp}$, $S_{headip}$, $S_{headpdcp}$, $G_k$, $K$, for $1\leq k\leq K$}
	\KwOut{$a_{k,n}, b_{k,j}$, for $1\leq k\leq K$, $1\leq j\leq q(N_k)$, $n\in \Omega$}
	\textbf{Initialize:} the total RB set $\Omega=\{1,...,N\}$ and the RBs determined to be assigned to client $k$, $N_k=\phi$, for $1\leq k\leq K$. Let $\mathcal{X}$ denote the set of all the clients, $\mathcal{X}=\{1,...K\}$ and $\mathcal{X}^{'}=\phi$. $a_{k,n}=0$, $b_{k,j}=0$, for $1\leq k\leq K$, $n\in \Omega$. Let $\Psi$ denote the set of the CQI values of all the RBs for all the clients, $\Psi=\{\varphi_{1,1},\cdots,\varphi_{1,N},\cdots,\varphi_{k,n},\cdots,\varphi_{K,N} \}$.
	
	\For{$k\in \mathcal{X}$}
	{
		Acquire the requirements of downlink transmission rate of client $k$, $DR_k=(\sum_{i=1}^{G_k}\frac{A}{T_{k,i}})/(G_k\cdot \alpha)$;
		
		Calculate the requirement of transmission rate for client $k$ at MAC layer, $M_k=DR_k(B_k+Nb_k(S_{headmac}+S_{headrlc}))/ P$;
		
		Obtain the transmission capacity of client $k$ in one scheduling period, $C_k=M_k\cdot TTI$;
	}
	\While{there are still some RBs left unassigned and the set of clients is not empty, $\Omega \neq \phi$ and $\mathcal{X}\neq \phi$}
	{
		\For{$k\in \mathcal{X}$}
		{
			Find $(k^*, n^*)$ with maximum CQI value,  $(k^{*},n^{*})=\underset{k,n}{arg\ max}\; \Psi$;
			
			Update the RB set of client $k^*$, $N_{k^*}=N_{k^*}\cup \{n^*\}$, and the RB assignment indicator $a_{k^*,n^*}=1$;
			
			Update the remaining available RB set $\Omega=\Omega\setminus \{n^*\}$. Determine
			the highest MCS that can support, $q(N_{k^*})$. Let $b_{k^*,j^*}=1$, if $j^*=q(N_{k^*})$, then calculate the transmission capacity of each RB, $r_{k^*,j^*}$;
			
			\If{client $k$'s transmission capacity requirement can be satisfied, $\sum_{n\in N_{k^*}}b_{k^*,j^*}r_{k^*,j^*}\geq C_k^{*}$}
			{
				The transmission capacity requirement of client $k^*$ can be
				satisfied, remove client $k^*$ into the set $\mathcal{X}^{'}$.
			}
		}
	}
	
	\While{there are still some RBs left unassigned and the set of clients is not empty, $\mathcal{X}^{'}\neq \phi $ and $\Omega\neq\phi$}
	{
		\For{$k\in \mathcal{X}^{'}$, $n\in \Omega$}
		{
			Find $(k^*, n^*)$ with maximum CQI value, $(k^{*},n^{*})=\underset{k,n}{arg\ max}\; \Psi$;
			
			Update the RB set of client $k^*$, $N_{k^*}=N_{k^*}\cup \{n^*\}$, and the RB assignment indicator $a_{k^*,n^*}=1$;
			
			Update the remaining available RB set $\Omega=\Omega\setminus \{n^*\}$. Determine
			the highest MCS that can support, $q(N_{k^*})$. Let $b_{k^*,j^*}=1$, if $j^*=q(N_{k^*})$, then calculate the transmission capacity of each RB, $r_{k^*,j^*}$;
			
			\If{$\sum_{n\in N_{k^*}}b_{k^*,j^*}r_{k^*,j^*}\geq J_k^{*}$}
			{
				The transmission capacity requirement of client $k^*$ can be
				satisfied, remove client $k^*$ from the set $\mathcal{X}^{'}$.
			}
		}
		
	}
\end{algorithm}

\subsection{TCP Congestion State and ACK Urgency Aware Resource Allocation for Uplink Video Transmission}
To keep smooth transmission, it is not enough to optimize the downlink video transmission because the blocked uplink feedback ACK will also deteriorate the congestion state of TCP, which in turn results in the degradation of downlink video transmission. To improve the feedback of ACK, we estimate the urgency of each ACK packet based on the TCP state information and RTO information of each client at eNodeB. Then, a TCP congestion state and ACK urgency aware uplink resource allocation scheme is performed by taking into consideration the ACK urgency, the TCP congestion state and the wireless network condition.

(1) Calculation of Uplink Scheduling Priority of Each ACK Packet

For uplink resource allocation at MAC layer of eNodeB, the urgency of each ACK packet is first considered because receiving feedback ACK timely is important to fluent transmission. Let $d_{k,\theta}$ be the time left for the transmission of ACK $\theta$ of client $k$ from client to TCP ACK receiving window. Then we can have
\begin{equation}\label{d_k,n}
d_{k,\theta}=RTO_k-(Tcur-Ts_{k,\theta})
\end{equation}
where $RTO_k$ is the maximum retransmission time interval threshold. $Tcur$ is the current system time and $Ts_{k,\theta}$ is the sending time of the corresponding TCP packet of ACK $\theta$ of client $k$.

Since only using ACK urgency information cannot reflect the comprehensive transmission state of client $k$, the congestion information such as the slow start threshold, $ssthresh_k$ , and the congestion window size, $cwnd_k$, are employed to determine the scheduling priority of each ACK packet. Here, we first add a modified weight $\eta_k$ to each client's RTO parameter $RTO_k$ to obtain the maximum allowed waiting time for ACK packets' arrival, $th_k=\eta_k\times RTO_k$, where $\eta_k$ can be regarded as the client playback experience within the range (0,1). The larger $\eta_k$  the worse playback experience. In this work, the playback experience is determined by the ratio of the times of the rebuffering events of client $k$ and that of all the clients. The scheduling priority of each ACK packet can be expressed by 
\begin{equation}\label{priority}\footnotesize
\left\{
\begin{aligned}
&sp_{k,\theta}=\frac{\delta}{d_{k,\theta}},\quad\quad\quad\;\;\quad\quad\quad\quad\; d_{k,\theta}<th_k\\
&sp_{k,\theta}=\frac{\beta}{ssthresh_k-cwnd_k}, \quad d_{k,\theta}\geq th_k,\; cwnd_k<ssthresh_k\\
&sp_{k,\theta}=\frac{\gamma}{cwnd_k-ssthresh_k}, \quad d_{k,\theta}\geq th_k,\; cwnd_k\geq ssthresh_k
\end{aligned}
\right.
\end{equation}
where $\delta$, $\beta$, $\gamma$ are the constant parameters which satisfy $\delta\gg \beta\gg \gamma$. From the Eq.(\ref{priority}), we can see that more urgent ACK packets will be given the highest scheduling priority. Once the time left for ACK packet becomes larger, the scheduling priority becomes lower. In addition, if the TCP congestion state of client $k$ is so terrible that $cwnd_k<ssthresh_k$, we will give its ACK packet higher scheduling priority correspondingly. 

(2) Problem Formulation

Before the uplink transmission problem formulation, we will first introduce the constraints that we need to comply with. Let $Z_k$ denote the number of the ACK packets which can be transmitted to client $k$ at the current scheduling period. It is assumed that all the ACK packets will be scheduled in FIFO (First In First Out) order. Thus we can have
\begin{equation}
\sum_{\theta=1}^{Z_k}\xi_{k,\theta}\leq \sum_{n\in \Omega}a_{k,n}\sum_{j=1}^{q(N_k)}b_{k,j}r_{k,j}<\sum_{\theta=1}^{Z_k+1}\xi_{k,\theta}
\end{equation}
where $\xi_{k,\theta}$ is the size of ACK packet $\theta$ of client $k$. Since $(a_{k,n}, b_{k,j}, r_{k,j})$ have the same meaning with the downlink transmission, they are also used in uplink transmission. Once $Z_k$ is obtained, we can easily acquire the final ACK scheduling result for each client as follows:
\begin{equation}
\tau_{k,\theta}=
\left\{
\begin{aligned}
&1,\quad 1\leq \theta\leq Z_k\\
&0,\quad Z_{k}+1\leq\theta\leq\widetilde{Z_k}
\end{aligned}
\right.
\end{equation}
where $\tau_{k,\theta}=1$ indicates ACK packet $\theta$ of client $k$ will be scheduled and $\tau_{k,\theta}=0$ otherwise. $\widetilde{Z_k}$ is the number of received ACK packets of client $k$. Therefore, the objective of our uplink transmission algorithm is to derive the value of $Z_k$.

The uplink transmission of ACK packets can be formulated into a mathematical model with the objective to maximum the sum of the scheduling priorities of the scheduled packets at the current scheduling period subject to the constraints of the uplink resource constraint and the packet scheduling strategy just as follows:
\begin{equation}\label{mubiao}
\begin{split}
&\mathbf{P}: max\sum_{k=1}^{K}\sum_{\theta=1}^{Z_k}sp_{k,\theta}\\
&s.t.\\ &\left\{
\begin{aligned}
&(c1)\;\sum_{\theta=1}^{Z_k}\xi_{k,\theta}\leq \sum_{n\in \Omega}a_{k,n}\sum_{j=1}^{q(N_k)}b_{k,j}r_{k,j} < \sum_{\theta=1}^{Z_k+1}\xi_{k,\theta}\\
&(c2)\;\sum_{k=1}^{K}a_{k,n}=1,a_{k,n}\in\{0,1\},n\in\Omega,k\in \mathcal{X}\\
&(c3)\;\sum_{j=1}^{q(N_k)}b_{k,j}=1,b_{k,j}\in \{0,1\},k\in \mathcal{X}\\
&(c4)\;\sum_{n\in\Omega}a_{k,n}\sum_{j=1}^{q(N_k)}b_{k,j}r_{k,j}\leq\sum_{\theta=1}^{\widetilde{Z_k}}\xi_{k,\theta}\\
&(c5)\;\tau_{k,\theta}\geq \tau_{k,\theta+1}
\end{aligned}
\right.
\end{split}
\end{equation}
Constraint (c4) and (c5) make sure that the transmission capacity cannot be greater than the sum size of the ACK packets of client $k$ and the ACK packets should be scheduled according to FIFO order respectively.

To solve the problem in Eq.(\ref{mubiao}), we introduce the utility function of RB $n$ for client $k$ as the improvement of the $\sum_{k=1}^{K}\sum_{\theta=1}^{Z_k}sp_{k,\theta}$ by assigning RB $n$ to client $k$ just as follows:
\begin{equation}
\begin{split}
L_{k,n}&=\max\left\{\sum_{\theta=1}^{Z_k}sp_{k,\theta}|\sum_{n^{'}\in N_k\cup\{n\}}a_{k,n^{'}}\sum_{j=1}^{q(N_k)}b_{k,j}r_{k,j}\right\}\\&-\max\left\{\sum_{\theta=1}^{Z_k}sp_{k,\theta}|\sum_{n^{'}\in N_k}a_{k,n^{'}}\sum_{j=1}^{q(N_k)}b_{k,j}r_{k,j}\right\}
\end{split}
\end{equation}
During the RB assignment, RB $n$ will be assigned to the client with maximum $L_{k,n}$.

(3) TU (TCP State and ACK Urgency Aware Uplink Resource Allocation) Algorithm

TU algorithm mainly consists of two steps: (1) Calculate the scheduling priority of all the ACK packets based on the information of the ACK urgency and the TCP congestion state. (2) Perform the RB assignment and MCS optimization for all the clients by using each RB's utility function. The detailed process of TU algorithm can be found in Algorithm \ref{uplink}.

\begin{algorithm}[!h]
	\caption{TU algorithm}
	\label{uplink}
	\KwIn{$Ts_{k,\theta}$, $Tcur$, $cwnd_k$, $ssthresh_k$, $RTO_k$, $Z_k$, $\widetilde{Z_k}$, $r_{k,j}$, $K$, $\eta_k$, $\delta$, $\beta$, $\gamma$, for $1\leq k\leq K$}
	\KwOut{$a_{k,n}, b_{k,j}$, for $1\leq k\leq K$, $1\leq j\leq q(N_k)$, $n\in \Omega$}
	\textbf{Initialize:} the total RB set $\Omega=\{1,...,N\}$ and the RBs determined to be assigned to client $k$, $N_k=\phi$, for $1\leq k\leq K$. Let $\mathcal{X}$ denote the set of all the clients, $\mathcal{X}=\{1,...K\}$. $a_{k,n}=0$, $b_{k,j}=0$, for $1\leq k\leq K$, $n\in \Omega$.
	\While{there are still some RBs left unassigned, $\Omega\neq\phi$}
	{
		\For{$n\in\Omega$ }
		{
			\For{$k\in \mathcal{X}$}
			{
				Update the RB set of client $k$, $N_k=N_k\cup\{n\}$ and the RB assignment indicator $a_{k,n}=1$;
				
				Obtain the optimal MCS selection and update the MCS indicator $b_{k,j^*}=1$;
				
				Acquire the time left for the transmission of ACK $\theta$ of client $k$, $d_{k,\theta}=RTO_k-(Tcur-Ts_{k,\theta})$;
				
				Calcultate the maximum allowed waiting time for ACK packets' arrival, $th_k=\eta_k\times RTO_k$;
				
				\eIf{$d_{k,\theta}<th_k$}
				{
					the scheduling priority of each ACK packet can be determined, $sp_{k,\theta}=\frac{\delta}{d_{k,\theta}}$;
				}
				{
					\eIf{$cwnd_k<ssthresh_k$}
					{
						$sp_{k,\theta}=\frac{\beta}{ssthresh_k-cwnd_k}$;
					}
					{	
						$sp_{k,\theta}=\frac{\gamma}{cwnd_k-ssthresh_k}$;
					}
				}
				Calculate the utility function of RB $n$ for client $k$, $L_{k,n}$;
				
			}
			Assign RB $n$ to client with maximum $L_{k,n}$, $k^{*}=\underset{1\leq k\leq K}{arg\ max}\; L_{k,n}$;
			
			Update other clients' RB set $N_k=N_k\setminus\{n\}$, RB assignment indicator $a_{k,n}=0$ and MCS indicator $b_{k,j^*}= 0$ for $k\neq k^*$.
			
			Update left unassigned RB set $\Omega=\Omega\setminus\{n\}$;
			
			\If{the transmission capacity is greater than the sum size of the ACK packets of client $k^*$, $\sum_{n\in\Omega}a_{k^*,n}\sum_{j=1}^{q(N_{k^*})}b_{k^*,j}r_{k^*,j}>\sum_{\theta=1}^{\widetilde{Z_{k^*}}}\xi_{k^*,\theta}$}
			{
				Remove client $k^*$ from the set $\mathcal{X}$;
			}
			
		}
		
	}
\end{algorithm}

\section{Results and Discussion}\label{result}

\subsection{Simulation configuration}
In this section, we present our experiment results carried out in LTE system with the NS3 (Network Simulation 3) simulator\cite{website:sim}. The specific system parameter configuration of LTE system is shown in Table \ref{tab:1}. The used standard video sequences are downloaded from the website\cite{website:video}. Each video sequence contains 8 video segments and each video segment consists of 60 frames. The playback frame rate is 30. The weights of all the inter-coded MBs for all the partition modes are set to be 0.2. The weights of intra-coded MBs with 4*4 and 16*16 partition modes in P frames are set to be 0.8 and 0.6 respectively. In the I frames, the weight of the 16*16 or the 4*4 intra-coded prediction mode is the same as that in the P frames. Detailed information of the requested videos can be found in Table \ref{tab:2}. To eliminate the impact of error concealment strategy on received video quality, we assume that error concealment for each user is the same. We extract the video encoding parameters to calculate the video packet importance level in the APP layer. Then, the video packet importance information is added to the TCP header in the transport layer. Finally, the information of TCP header is inspected in MAC layer. To examine the performance of our proposed algorithms, we employ different combinations of the following state-of-art algorithms for uplink and downlink transmission.

\begin{table}[!h]
	\centering
	\caption{LTE Configuration}
	\label{tab:1}
	\begin{tabularx}{8cm}{|p{4cm}<{\centering}|p{3.3cm}<{\centering}|}
		\hline
		\textbf{Parameters} & \textbf{Values}\\
		\hline
		Number of Downlink RBs & 16,18,20,22\\
		\hline
		Number of Uplink RBs &  8\\
		\hline
		Subcarrier per RB &  12\\
		\hline
		Subcarrier Spacing & 15 KHz \\	
		\hline
		Bandwidth per RB & 180 KHz\\
		\hline
		eNodeB Tx Power & 43 dBm\\
		\hline
		Transmission Time Interval & 1 ms\\
		\hline
		Carrier Frequency & 2 GHz \\
		\hline
		Channel Model & Typical Urban\\
		\hline
		Propagation Model & Macro-Cell Urban\\
		\hline
		Path Loss & 20 dB\\
		\hline
		Modulation & QPSK,16QAM,64QAM \\
		\hline
		Turbo Coding Rate & 1/2, 3/4\\
		\hline
		CQI Report & Full Bandwidth\\
		\hline
		CQI Report Period & 1 ms \\
		\hline
		CQI Rangehh & [1,7], [8,15]\\
		\hline
	\end{tabularx}
	
\end{table}

\begin{table*}[!h]
	\centering
	\caption{Video Configuration}
	\label{tab:2}
	\begin{tabularx}{15.3cm}{|p{2.25cm}<{\centering}|p{1.25cm}<{\centering}|p{1.25cm}<{\centering}|p{1.25cm}<{\centering}|p{1.25cm}<{\centering}|p{1.25cm}<{\centering}|p{1.25cm}<{\centering}|p{1.25cm}<{\centering}|p{1.25cm}<{\centering}|}
		\hline
		\diagbox[width=7.9em,height=6em]{Video Sequence}{Coding Rate} & \textbf{Segment 1 (kbps)}&\textbf{Segment 2 (kbps)}&\textbf{Segment 3 (kbps)}&\textbf{Segment 4 (kbps)}&\textbf{Segment 5 (kbps)}&\textbf{Segment 6 (kbps)}&\textbf{Segment 7 (kbps)}&\textbf{Segment 8 (kbps)}\\
		\hline
		\textbf{flower.cif} & 1509.4 & 1784.2& 1642.9 & 1945.8 & 1615.4 & 1718.9 & 1658.0 & 1895.6\\
		\hline
		\textbf{coastguard.cif} &  1316.7& 1302.7 & 1051.4 & 1000.2 & 945.9 & 1316.7 & 1302.7 & 1051.4\\
		\hline
		\textbf{news.cif} & 333.0 & 387.5 & 362.4 & 356.6 & 370.3 & 333.0 & 387.5 & 362.4 \\
		\hline
		\textbf{highway.cif} & 250.1 & 286.9 & 258.6 & 265.6 & 273.1 & 252.9 &269.5 & 370.2\\
		\hline
		\textbf{soccer.cif} &  744.1 & 816.9 & 913.2 & 773.8 & 899.8 & 744.1 & 816.9 & 913.2\\
		\hline
		\textbf{foreman.cif}& 470.6 & 464.8 & 589.3 & 646.6 & 600.8 & 470.6 & 464.8 & 589.3\\
		\hline
		\textbf{crew.cif} & 618.2 & 965.8 & 1189.0 & 918.5 &925.1 & 618.2 & 965.8 &1189.0\\
		\hline
		\textbf{bus.cif} & 1470.2 & 1351.8 & 1439.6 & 1390.3 & 1405.3 & 1470.2 & 1351.8 & 1439.6 \\	
		\hline
	\end{tabularx}
	\label{tab:video_configuration}
\end{table*}

\vspace{0.3cm}\textbf{(i) PF (Proportional Fair) cscheduling algorithm}

\vspace{0.3cm}
PF algorithm \cite{PF} refers to the resource allocation scheme which allocates scheduling priority to clients based on the current channel states and the previous client throughput, to maximize the system throughput and guarantee fairness among the clients.

\vspace{0.3cm}\textbf{(ii) RR (Round-Robin) scheduling algorithm}

\vspace{0.3cm}
RR algorithm \cite{RR} provided a fair resource allocation scheme, RR, at MAC layer without considering different requirements among clients at the APP layer. The resource was just allocated to the clients in a round-robin mode.

\vspace{0.3cm}\textbf{(iii) MAXCI (Maximum Carrier Interference) scheduling algorithm}

\vspace{0.3cm}
MAXCI algorithm \cite{Maxci} developed a purely channel-dependent scheduling scheme, MAXCI, where the clients with good channel states always occupy the channel most of time even if they are well-ahead of their video decoding deadlines.

\vspace{0.3cm}\textbf{(iv) MLWDF (Modified Largest Weighted Delay First) scheduling algorithm}

\vspace{0.3cm}
In the MLWDF scheduling algorithm \cite{MLWDF}, the scheduler prioritized clients with the delays of their head-of-line packets and channel states. In other words, it comprehensively considers the channel throughput and data caching queue, to achieve the optimal scheduling scheme.

Note that TU$\_$TD denotes TU algorithm for uplink transmission and TD algorithm for downlink transmission respectively, and APD$\_$TU$\_$TD denotes the TU$\_$TD combined with APD strategy. This rule applies to the following mentioned abbreviations. 

\subsection{Performance Evaluation}
To evaluate the efficiency of our algorithm, we compare our proposed resource allocation algorithm with the existing classical ones, mainly considering the system throughput, TCP window size, the rebuffering time and the PSNR of the clients. 

\subsubsection{Performance Evaluation of TU$\_$TD}

In this subsection, the numbers of the clients, uplink RBs and downlink RBs are 8, 8 and 16, respectively. The aim is to evaluate the performance of the system throughput, the TCP window size and the rebuffering time for different algorithms. 

\begin{figure*}[htb]
	\centering
	\includegraphics[width=17cm]{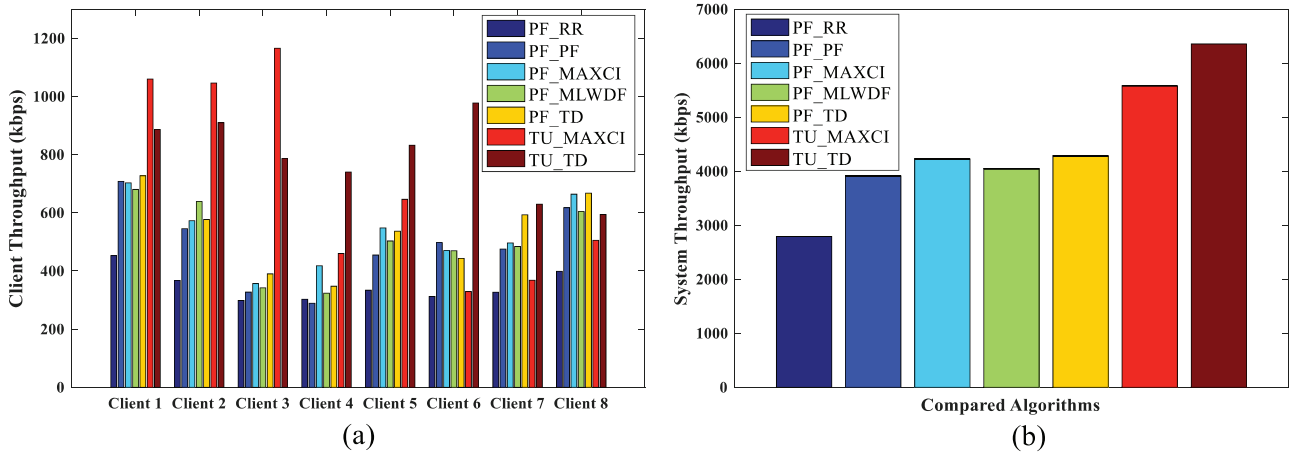}
	\caption{(a) The throughput of each client. (b) The system throughput of different algorithms.}
	\label{fig:tuntuliang}
\end{figure*}

As can be seen from Fig. \ref{fig:tuntuliang} (a), the throughput of most of the clients for PF, RR, MAXCI and MLWDF cannot match the bit rate of the requested video as shown in Table \ref{tab:video_configuration}. In this case, the rebuffering events will occur. In contract, the proposed TU$\_$TD algorithm can greatly improve the clients' throughput and meet the demand of transmission rate of most clients. For client 1 to 3, the throughput of the proposed joint optimization algorithm TU$\_$TD is not higher than that of TU$\_$MAXCI. This is because the TU$\_$MAXCI prefers the clients with good channel conditions and allocates the resources to them. Although the throughput of some clients increases, the performance of other clients is sacrificed. This results in a low transmission capacity for the clients with poor channel state, which will affect the system throughput of TU$\_$MAXCI. As shown in Fig. \ref{fig:tuntuliang} (b), the TU$\_$TD algorithm performs best in the respective of system throughput and obtains 13.9$\%$ improvement compared with TU$\_$MAXCI. For other algorithms, TU$\_$TD can acquire at least 50.3$\%$ system throughput improvement. For example, the system throughput of TU$\_$MAXCI and TU$\_$TD are 5580.3 kbps and 6356.2 kbps respectively. From the above observations, we can say that our proposed TU$\_$TD can improve the system throughput significantly. It is because the ACK feedback rate is considered to estimate the transmission capacity of MAC layer in TD algorithm and ACK urgency is utilized to guide uplink resource allocation, which can keep the balance of the ACK feedback rate and downlink resource allocation. This can decrease the possibility of the TCP window size degradation due to the packet loss caused by unreasonable resource allocation. 

\begin{figure*}[!h]
	\centering
	\includegraphics[width=17cm]{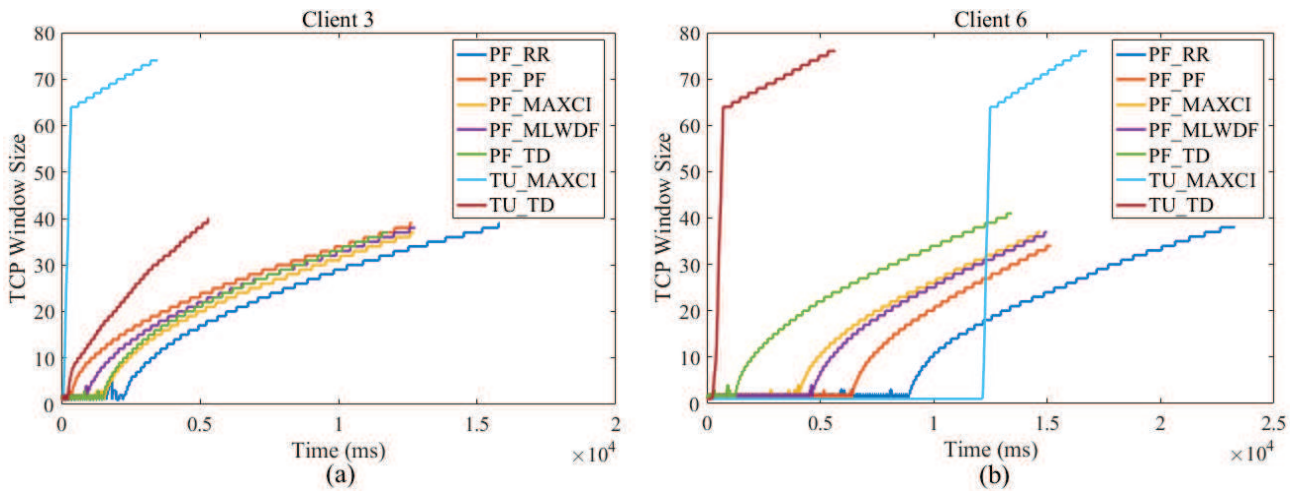}
	\caption{(a) TCP window size of client 3 who requests news.cif (b) TCP window size of client 6 who requests foreman.cif.}
	\label{fig:TCP}
\end{figure*}

Fig. \ref{fig:TCP} (a) shows the results of the change of TCP window size for all the algorithms. We can see that, in the initial stage, the TCP window size of client 3 is always halved. With the improvement of channel state, its TCP window size keeps increasing. From the respective of the whole transmission process, the proposed TU$\_$MAXCI algorithm performs best, and only employs 4471 ms to complete the video transmission of client 3. This is because the channel state of client 3 is superior to other clients. The finish of the transmission for client 3 can leave more resources to other clients who are really in need. PF$\_$ RR algorithm performs the worst and the window size of the client 3 remains unchanged for a long while, which results in a longer initial startup delay. In Fig. \ref{fig:TCP} (b), we can observe that except for TU$\_$TD algorithm, other algorithms suffer many timeout retransmissions at the initial stage, which will cause the occur of the rebuffering events during the initial playback. Fortunately, the TCP window size of the proposed joint optimization algorithm TU$\_$TD increases rapidly after a short time, which greatly improves the throughput of client 6. Therefore, the total transmission time of the requested video for client 6 is fewer than that of other compared algorithms, which in turn validates our proposed optimization algorithm TU$\_$TD.

\begin{figure*}[!h]
	\centering
	\includegraphics[width=17.5cm]{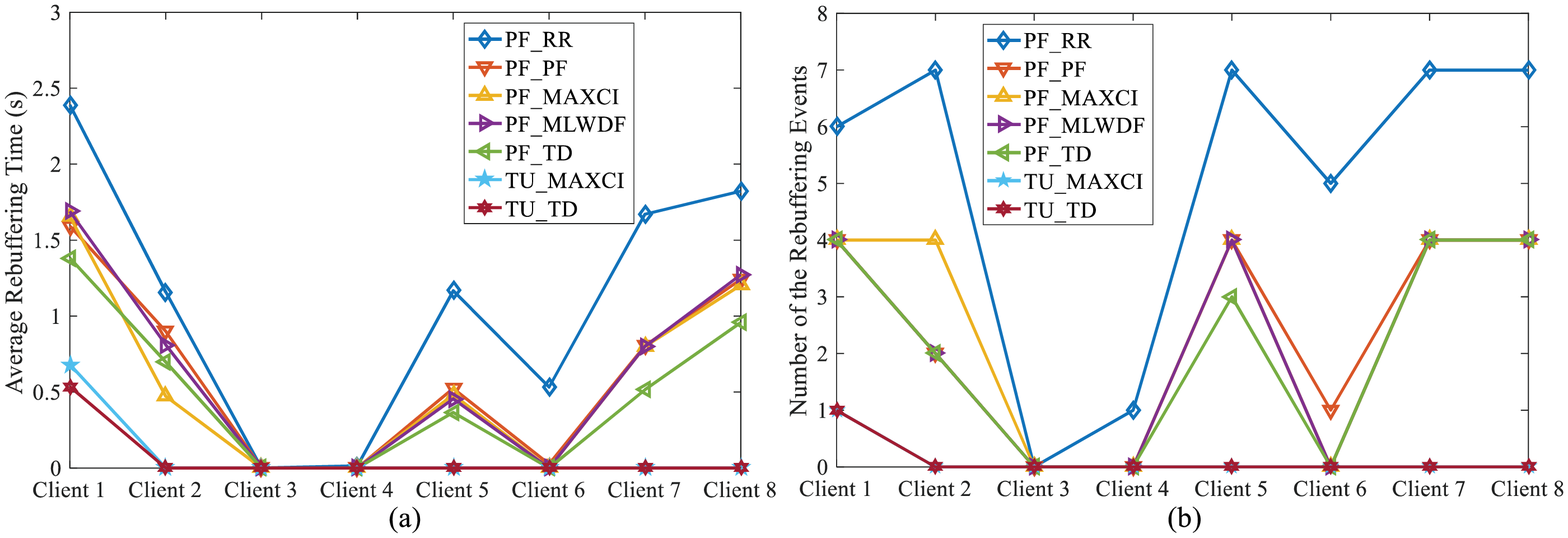}
	\caption{(a) The average rebuffering time of each client (b) The number of the rebuffering events of each client.}
	\label{fig:bofangkadun}
\end{figure*}

Fig. \ref{fig:bofangkadun} (a) and Fig. \ref{fig:bofangkadun} (b) show the average rebuffering time and the number of rebuffering events of 8 clients for different algorithms respectively. We can see that The TU$\_$TD and TU$\_$MAXCI can provide continuous playback experience for most clients except for client 1. The total rebuffering time of client 1 for TU$\_$MAXCI is 27.2$\%$ longer than TU$\_$TD and the number of the rebuffering events of TU$\_$TD is the same with that of TU$\_$MAXCI. For these two algorithms, only client 1 suffers one rebuffering event during the whole playback process. In addition, it can also be seen that the client 3 never encounters the playback rebuffering event for all the compared algorithms. This is because the bit rate of the requested video of client 3 is relatively lower than other videos and such that the continuous playback can be easily guaranteed even if the uplink and downlink resource allocation schemes are unreasonable. Based on the results shown in Fig. \ref{fig:tuntuliang} and Fig. \ref{fig:bofangkadun}, we can say that the playback experience is related to the system throughput and improving the system throughput can bring more satisfactory playback experience.

\subsubsection{Performance Evaluation of APD}
To better validate the performance of our proposed autonomous packet drop strategy, we carry out some simulations by combing APD with TU and TD. To measure the effect of the packet loss on the playback experience and throughput, the channel state is kept poor in the simulation of this subsection.
\begin{figure}[!h]
	\centering
	\includegraphics[width=8cm]{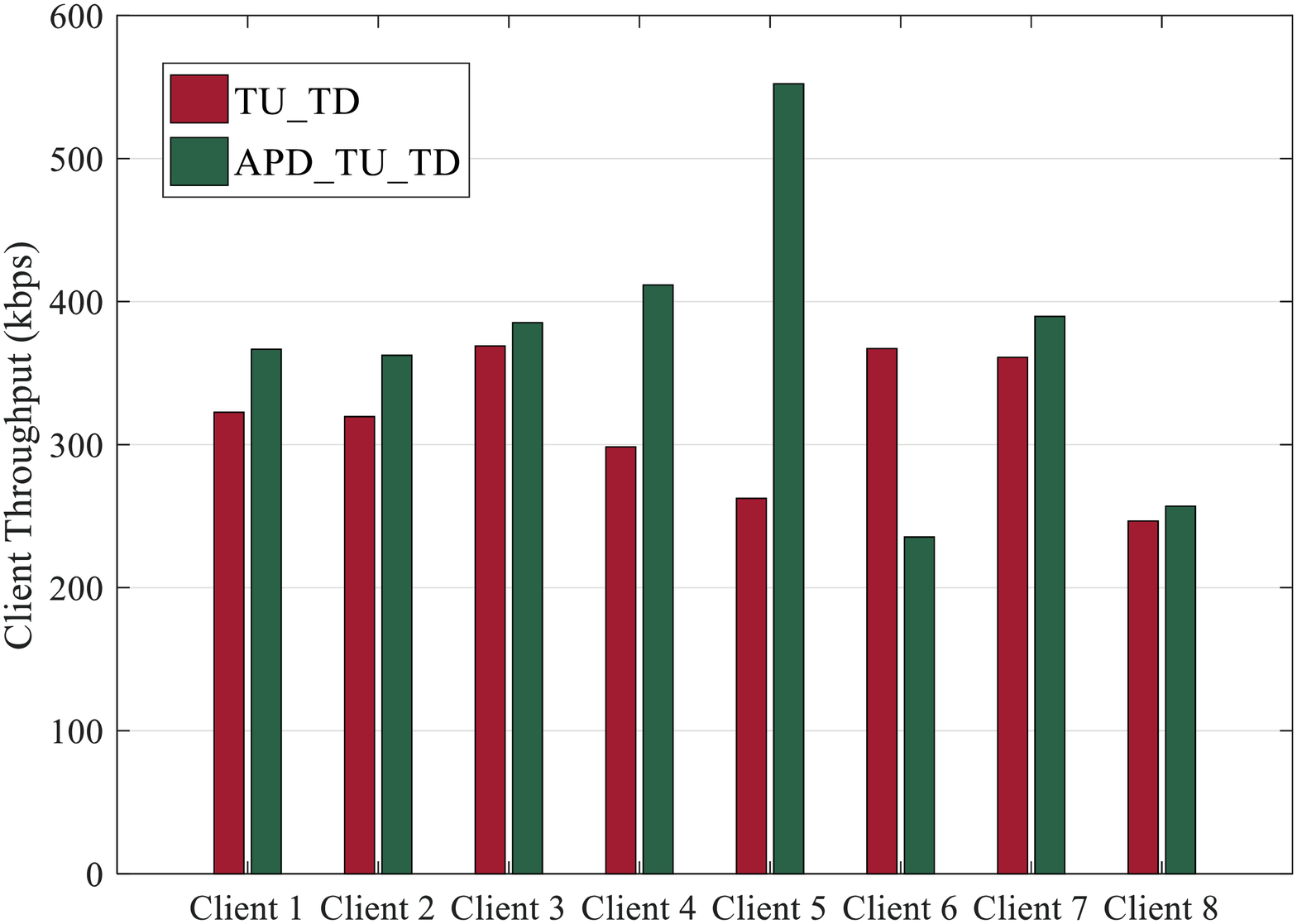}
	\caption{The throughput of each client.}
	\label{fig:tuntuliang_cha}
\end{figure}

From Fig. \ref{fig:tuntuliang_cha}, it can be seen that under poor channel state, APD$\_$TU$\_$TD can acquire higher throughput than TU$\_$TD for most of the clients. But obviously the system throughput of APD$\_$TU$\_$TD is higher than that of TU$\_$TD. Statistical results show that APD$\_$TU$\_$TD algorithm can acquire 16.14$\%$ improvement over TU$\_$TD algorithm. This is because the possibility of the TCP window size degradation caused by packet loss of unimportant packets can be efficiently decreased by discarding unimportant video packets before putting them into the TCP sending window. This will in turn improve the throughput of each client and the system throughput. 
\begin{figure}[!h]
	\centering
	\includegraphics[width=8cm]{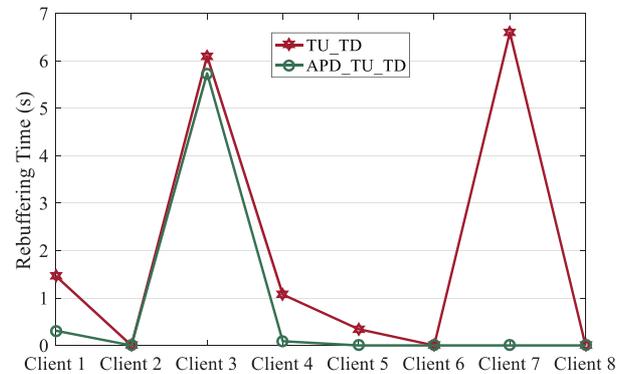}
	\caption{The rebuffering time of each client.}
	\label{fig:huanchong_bad}
\end{figure}

\begin{table*}[!h]
	\centering
	\caption{The performance in terms of PSNR and rebuffering time for APD$\_$TU$\_$TD and TU$\_$TD}
	\begin{tabularx}{16cm}{|p{1.9cm}<{\centering}|p{1.3cm}<{\centering}|p{1.5cm}<{\centering}|p{1.9cm}<{\centering}|p{1.9cm}<{\centering}|p{1.9cm}<{\centering}|p{1.2cm}<{\centering}|p{1.7cm}<{\centering}|}
		\hline
		Video Sequence                       & Client Index        & TU\_TD PSNR (dB)     & APD\_TU\_TD PSNR (dB) & TU\_TD Rebuffering Time (s) & APD\_TU\_TD Rebuffering Time (s) & PSNR Decline (\%) & Rebuffering Time Decline (\%) \\ \hline
		\multirow{2}{*}{news.cif}            & 1                   & \multirow{2}{*}{38.97} & 38.01                 & 1.466                       & 0.31                             & 2.48              & 78.76                         \\ 
		& 5                   &                        & 38.32                 & 0.344                       & 0                                & 1.66              & 100                           \\ \hline
		\multirow{2}{*}{highway.cif}         & 2                   & \multirow{2}{*}{38.50} & 38.36                 & 0                           & 0                                & 0.36              & 0                             \\
		& 6                   &                        & 38.06                 & 0                           & 0                                & 1.14              & 0                             \\ \hline
		\multirow{2}{*}{crew.cif}            & 3                   & \multirow{2}{*}{37.49} & 36.77                 & 6.098                       & 5.736                            & 1.92              & 5.93                          \\ 
		& 7                   &                        & 36.91                 & 6.601                       & 0                                & 1.54              & 100                           \\ \hline
		\multirow{2}{*}{foreman.cif}         & 4                   & \multirow{2}{*}{37.03} & 36.50                 & 1.073                       & 0.089                            & 1.43              & 91.71                         \\
		& 8                   &                        & 36.46                 & 0                           & 0                                & 1.53              & 0                             \\ \hline
		\multicolumn{2}{|c|}{\makecell[c]{Average PSNR\\ Performance}}             & 37.99                  & 37.42                 & /                           & /                                & 1.50              & /                             \\ \hline
		\multicolumn{2}{|c|}{\makecell*[c]{Average Rebuffering \\ Time Performance}} & /                      & /                     & 1.94                        & 0.76                             & /                 & 60.82                         \\ \hline
	\end{tabularx}
	\label{tab:psnr}
\end{table*}

As can be seen from Fig. \ref{fig:huanchong_bad}, for APD$\_$TU$\_$TD, the clients experience shorter rebuffering time than that of TU$\_$TD particularly client 7. Client 2, 5, 6, 7 and 8 can always keep continuous playback by combining APD with TU and TD. The client 3 suffers a long rebuffering time because the requested crew.cif has the highest rate and the channel state is unable to support timely transmission of video packets even though more RBs are provided. Since the rebuffering time is the most important factor which affects clients' QoE most, APD$\_$TU$\_$TD can acquire better playback experience than TU$\_$TD. It is easy to understand that the smaller number of the video packets to be transmitted the lower probability of the TCP window size degradation due to the packet loss. Larger transmission capacity and less video packets to be transmitted can avoid some rebuffering events. Motivated by this, some unimportant video packets are dropped autonomously by our proposed packet drop strategy before they are putted into TCP sending window.

Table \ref{tab:psnr} shows the performance in terms of PSNR and rebuffering time for APD$\_$TU$\_$TD and TU$\_$TD. One observation is that PSNR of APD$\_$TU$\_$TD is slightly lower than that of TU$\_$TD for all the clients. This is because APD$\_$TU$\_$TD will drop some unimportant packets to adjust the amount of the video packets that TCP should afford according to the information of the playback status of client and the importance derived from the encoder. Since our proposed algorithm APD only discards the video packets with small size and low importance, the reduction of PSNR is so small that its impact on the clients' playback experience is negligible. Statistical results show that the average PSNR decline of all the videos is just about 1.50$\%$. The other observation is that the performance of APD$\_$TU$\_$TD is superior to TU$\_$TD in terms of rebuffering time. Statistical results show that APD$\_$TU$\_$TD can achieve 60.82$\%$ improvement of the average rebuffering time of all the clients. In addition, more clients can enjoy continuous playback experience by using APD$\_$TU$\_$TD. As the most important factor of QoE, such significant degradation of rebuffering time efficiently optimizes the clients' playback experience.

\begin{figure*}[htb]
	\centering
	\includegraphics[width=17cm]{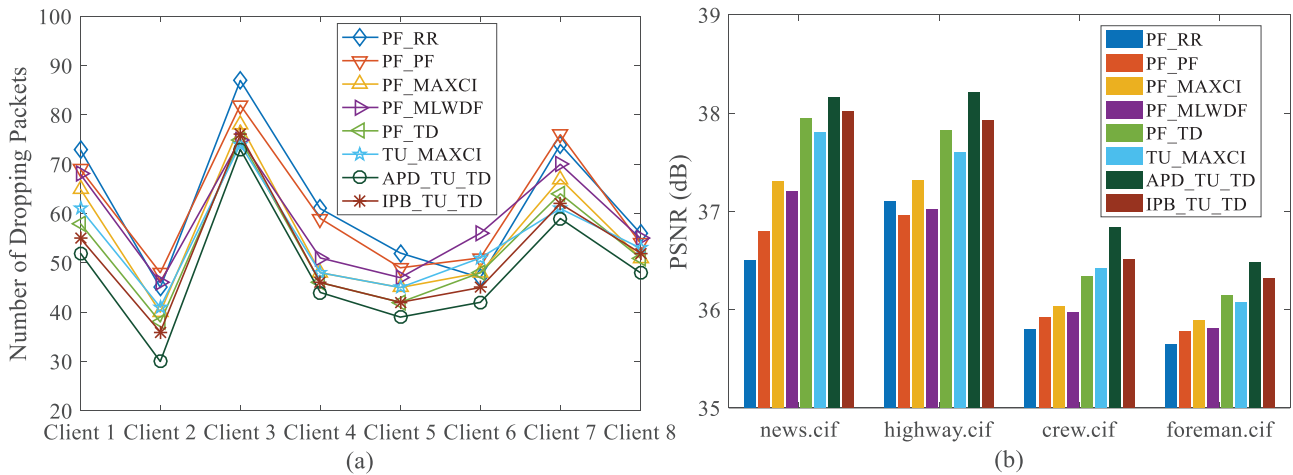}
	\caption{(a) Number of dropping packets of each client for different algorithms. (b) PSNR of different video sequences with dropping packets for different algorithms. The number of video sequence packets (news.cif, highway.cif, crew.cif, foreman.cif) are 1042, 805, 2276 and 1432, respectively. }
	\label{fig:packet_PSNR}
\end{figure*}

Fig. \ref{fig:packet_PSNR} (a) and Fig. \ref{fig:packet_PSNR} (b) show the number of dropped packets of each client and the PSNR of four video sequences for different algorithms, respectively. IPB$\_$TU$\_$TD is formulated by combing our proposed TU$\_$TD method with the dropping strategy of different priorities based on I, P, B frames. It can be seen that our proposed APD$\_$TU$\_$TD algorithms outperforms other compared algorithms in terms of dropping packets and PSNR, which validates the effectiveness of our proposed algorithm.

\subsubsection{Performance Evaluation of Different Number of the Clients and RBs}
\begin{figure}[!h]	
	\centering
	\includegraphics[width=8.5cm]{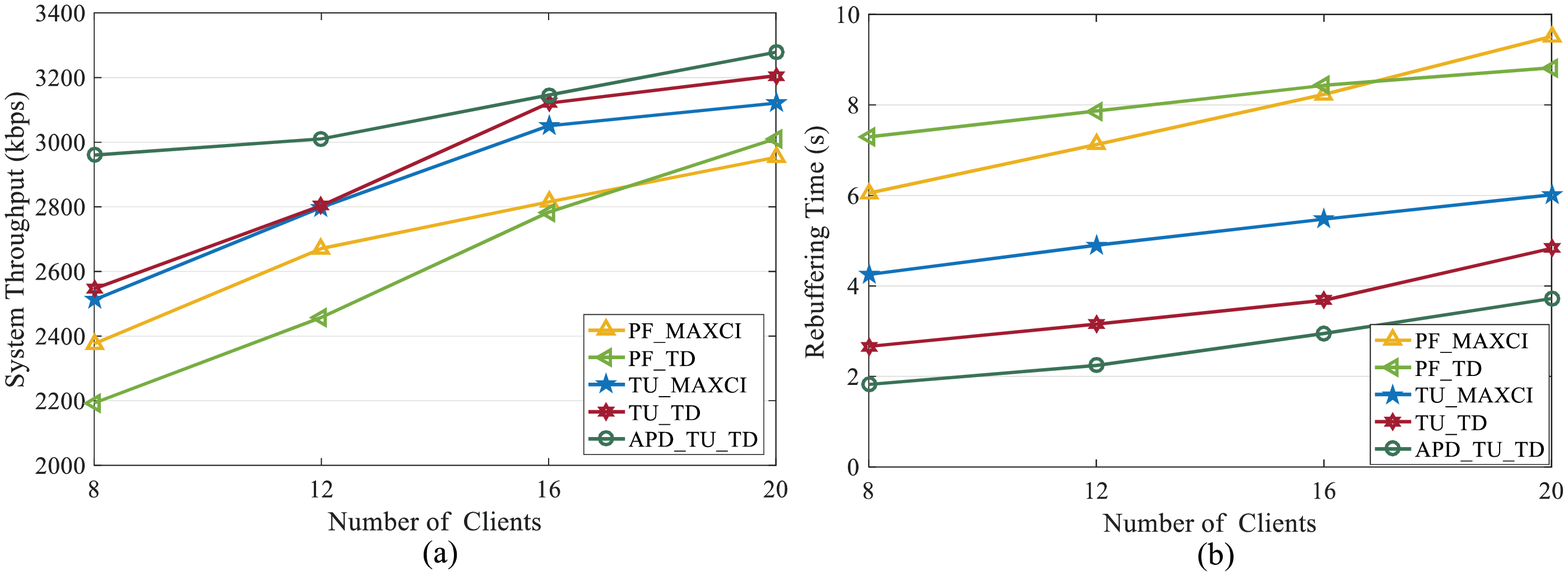}
	\caption{(a) The system throughput vs. the number of clients. (b) The rebuffering time vs. the number of clients.}
	\label{fig:tuntu_bofang}
\end{figure}

Fig. \ref{fig:tuntu_bofang} (a) and Fig. \ref{fig:tuntu_bofang} (b) show the system throughput and rebuffering time with different numbers of the clients respectively. We can see that both the system throughput and the rebuffering time become higher with the increase of the number of RBs. It is easy to understand that more clients will introduce more competition on the limited wireless resource, which will improve the efficiency of the resource usage. But we can also see that the improvement of system throughput caused by competition tends to be stable when the number of the clients becomes 20. This is because the resource utilization efficiency has reached its upper bound. In addition, it can be seen that our proposed APD$\_$TU$\_$TD algorithm outperforms other compared algorithms in terms of system throughput as well as the rebuffering time for different numbers of the clients. TU$\_$TD algorithm can acquire close performance to APD$\_$TU$\_$TD. Overall, we can say that considering the information of the APP layer, the transport layer and the MAC layer can improve the system throughput and the playback experience of the clients.

\begin{figure}[!h]
	\centering
	\includegraphics[width=8.5cm]{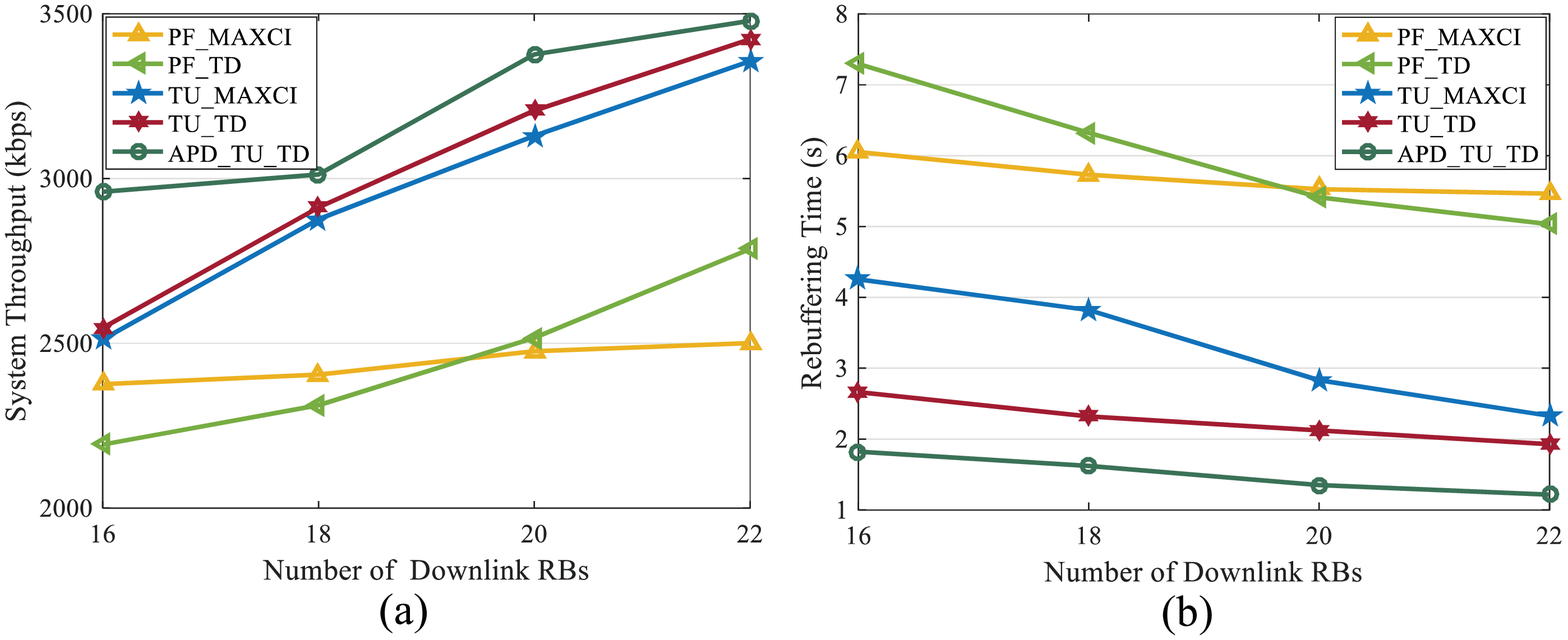}
	\caption{(a) The system throughput vs. the number of downlink RBs. (b) The rebuffering time vs. the number of downlink RBs.}
	\label{fig:RBzengjia}
\end{figure}

We also examine the performance of system throughput and rebuffering time with different numbers of RBs in Fig. \ref{fig:RBzengjia}. We can see that the system throughput becomes higher and rebuffering time becomes shorter with the increase of the number of RBs. It is easy to understand that more RBs can provide higher transmission capacity and therefore can perform better system throughput and more continuous playback. In addition, it can be seen that for different numbers of the RBs, the system throughput of our proposed APD$\_$TU$\_$TD algorithm is always higher than that of TU$\_$TD algorithm. This indicates that packet loss will occur even if the number of the RBs becomes larger. Our proposed APD algorithm can reduce the waiting time which affords the transmission of some unimportant packets. When the number of downlink RBs is 22, APD$\_$TU$\_$TD can earn 39.08$\%$ system throughput improvement compared with PF$\_$MAXCI. For rebuffering time, APD$\_$TU$\_$TD can achieve almost continuous playback while PF$\_$MAXCI will bring 5.4 s playback interruption to the clients.

\begin{figure}[!h]
	\centering
	\includegraphics[width=8.5cm]{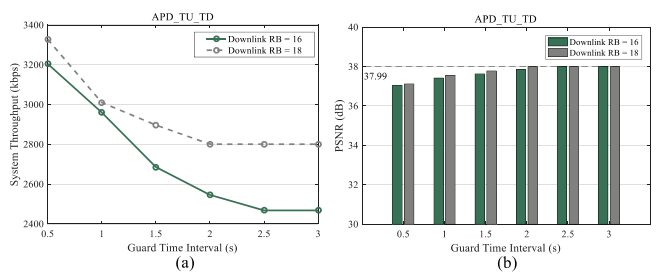}
	\caption{(a) The system throughput vs. the guard time interval with APD$\_$TU$\_$TD algorithm. (b) The average PSNR of requested videos vs. the guard time interval with APD$\_$TU$\_$TD algorithm.}
	\label{fig:baohu_psnr}
\end{figure}

Finally, we provide the performance of the system throughput and the average PSNR of requested videos with various guard time intervals and different numbers of downlink RBs in Fig. \ref{fig:baohu_psnr}. We can see that when the guard time interval becomes longer, the system throughput gradually decreases and PSNR gradually increases. This is because with the increase of guard time interval, the chance left for APD becomes slim and therefore the number of unimportant packets, which need to be dropped, will also becomes less. The limited wireless resource has to afford the transmission of the useless packets, whose packet loss may lead to degradation of TCP window size. Besides, the client needs to spend more time on the arrival of these packets. When the guard time interval exceeds 2.5 s, the system throughput and PSNR become stable because no autonomous packet loss occurs. We can also see that with more downlink RBs, the system throughput becomes less sensitive to APD. The reason is that the probability of packet loss in a scenario with sufficient downlink resource is lower than that with limited resource. Therefore, the probability of TCP window size degradation due to the packet loss becomes lower, which reduces the effect of APD, in a scenario with more downlink RBs.

\section{Conclusion}\label{conclusion}
This paper considered a playback experience driven cross layer design of the APP layer, the transport layer and the MAC layer to improve the system throughput as well as the playback experience of the video clients in LTE system. At the transport layer, a QoE requirement aware autonomous packet drop strategy is developed to keep a balance between the video packets to be transmitted and the current wireless network condition based on the information of packet importance derived from the encoder and the channel state. Then, feedback ACK information is used to calculate the transmission capacity requirement to guide the downlink resource allocation at the MAC layer. Our proposed TCP state aware downlink resource allocation scheme can avoid allocating excessive resources to the clients who don't really need. Besides, to improve the uplink feedback of ACK, we further develop a packet urgency-based resource allocation strategy based on RTO and TCP congestion window information. Simulation results showed that joint optimization of the APP layer, the transport layer and the MAC layer can achieve superior system throughput, less rebuffering events and acceptable PSNR simultaneously. 

\section{Acknowledgments}\label{sec11}

This research was supported in part by the National Science Foundation of China(61701389, U1903213), the Natural Science Basic Research Plan in Shaanxi Province of China(2018JQ6022) and Shaanxi Key R\&D Program(2018ZDCXL-GY-04-03-02).

\end{document}